\documentclass[prl,twocolumn,showpacs,preprintnumbers,amsmath,amssymb,floatfix,aps]{revtex4-1}

\usepackage{graphicx}
\usepackage{dcolumn}
\usepackage{bm}

\begin{document}

\title{Collision of two spin polarized fermionic clouds}

\author{O. Goulko}
\affiliation{Department of Applied Mathematics and Theoretical Physics, University of Cambridge, Centre for Mathematical Sciences, Cambridge, CB3 0WA, United Kingdom}
\email{O.Goulko@damtp.cam.ac.uk}
\author{F. Chevy}
\affiliation{Laboratoire Kastler Brossel, Ecole Normale Sup\'erieure, 24 rue Lhomond, 75231 Paris Cedex 05, France}
\author{C. Lobo}
\affiliation{School of Mathematics, University of Southampton, Highfield, Southampton, SO17 1BJ, United Kingdom}


\begin{abstract}
We study the collision of two spin polarized Fermi clouds in a harmonic trap using a simulation of the Boltzmann equation. As observed in recent experiments we find three distinct regimes of behavior. For weak interactions the clouds pass through each other. If interactions are increased they approach each other exponentially and for strong interactions they bounce off each other several times. We thereby demonstrate that all these phenomena can be reproduced using a semiclassical collisional approach and that these changes in behavior are associated with an increasing collision rate. We then show that the oscillation of the clouds in the bounce regime is an example of an unusual case in quantum gases: a nonlinear coupling between collective modes, namely the spin dipole mode and the axial breathing mode which is enforced by collisions. We also determine the frequency of the bounce as a function of the final temperature of the equilibrated system.
\end{abstract}
\pacs{03.75.Ss, 03.75.Hh}

\maketitle
Spin transport has become in recent years a much studied field in solid state systems, for example in mesoscopic phenomena, where the electronic spin degree of freedom is used to create new devices. Understanding the spin relaxation, diffusion and other transport properties is of fundamental importance in such fields. In atomic gases it has been studied mainly in spinor Bose gases \cite{Bose1, Bose2}. There has been a renewed interest in spin transport in Fermi gases which are a clearer parallel to electronic systems \cite{du2008observation, du2009controlling, piechon2009large, Duine, Zwierlein, Bruun}. An important advantage of cold gases in such studies is the simplification due to the absence of relaxation mechanisms for spin currents apart from direct collisions between atoms of different spin, unlike e.g.\ in a solid where collisions with the ionic lattice can be important. In addition, the atomic interaction and initial temperature of the clouds are easily tuneable parameters. Finally, as we shall see, very large spin polarizations can be easily created leading to large spin currents.

Here we study the collision of two clouds with opposite spin polarization following the recent experiments of A.~Sommer et al.~\cite{Zwierlein}. We confine ourselves to the semiclassical regime, using a Boltzmann equation simulation. In contrast, a recent theoretical study has instead used a hydrodynamic approach based on a many-body equation of state \cite{Taylor2011Colliding}.

One of the most striking experimental observations was the bouncing of the clouds off each other. We will here demonstrate that this phenomenon can be understood purely in terms of semiclassical collisions, without recourse to e.g.\ mean fields or other more complicated effects. For instance, our approach predicts that all quantities considered here depend only on the square of the scattering length and not on its sign. Also, as we will show, the bouncing oscillations can be understood as an example of a nonlinear coupling between collective modes which, to our knowledge, has never before been studied in Fermi gases \footnote{In Bose gases, nonlinear coupling between modes can lead to damping of collective excitations, as in the case of Landau and Beliaev damping.}.

We consider a system of two-component fermions labelled by the spin index $s=\{\uparrow,\downarrow\}$ with equal mass $m$. We set $\hbar=k_B=1$ throughout. Fermions of opposite spin can interact via $s$-wave collisions and the cross section is given by $\sigma=4\pi a^2/(1+a^2\mathbf{p}_\textnormal{rel}^2/4)$, where $a$ is the scattering length and $\mathbf{p}_\textnormal{rel}$ the relative momentum of the two atoms. The fermions are confined in a cigar shaped harmonic trap with potential $V(\mathbf{r})=\frac{1}{2}m(\omega_x^2x^2+\omega_y^2y^2+\omega_z^2z^2)$, where $\omega_z<\omega_x=\omega_y$.

We assume that the system is in the normal phase and that the temperature is sufficiently high, so that the two spin distributions can be described semiclassically in terms of functions $f_s(\mathbf{r},\mathbf{p},t)$. The initial distributions are created by sampling a Fermi-Dirac distribution $f(t=0)=(e^{(p^2/2m+V(\mathbf{r})-\mu)/T_{\rm init}}+1)^{-1}$ normalized to $N/2=N_\uparrow=N_\downarrow$, where $T_{\rm init}$ is the temperature and $\mu$ the chemical potential. The atom number $N_s$ for each species is held fixed during the simulation. Then each spin distribution is displaced in opposite directions along the $z$-axis by $d_0/2$. In the subsequent time evolution, the clouds begin to move towards the center under the harmonic trapping force resulting in a collision between them. After a sufficiently long time the center of mass energy will be transformed into the internal energy of the gas and a new equilibrium state will be reached, characterized by a Fermi-Dirac distribution with temperature $T_{\rm final}$ (and a chemical potential $\mu_{\rm final}$). Note that $T_{\rm final}$ is a function of only the atom number, $T_{\rm init}$ and $d_0$ and can be calculated exactly from these values using energy conservation.

Simulations were carried out for $N=10000$, a range of initial cloud temperatures $0.2\leq T_{\rm init}/\tilde{T}_F\leq10$, interaction strength $0.5\leq|\tilde{k}_Fa|\leq10$, and initial distances between the centers of mass of the two clouds $0.4\sigma_z\leq d_0\leq16\sigma_z$, where $\tilde{T}_F=\tilde{k}_F^2/2m=(3N\omega_x\omega_y\omega_z)^{1/3}$ and $\sigma_z=\sqrt{T_\textnormal{init}/m\omega_z^2}$. Our numerical setup is similar to the one described in \cite{urban}. Here we will only give a brief summary with a more detailed paper to follow \cite{inprep}.

The time-evolution of the distribution function $f_s(\mathbf{r},\mathbf{p},t)$ is given by the Boltzmann equation,
\begin{equation}
\partial_t f_s+(\mathbf{p}/m)\cdot\nabla_rf_s-\nabla_rV\cdot\nabla_pf_s=-I[f_s,f_{\overline{s}}],
\end{equation}
where the left-hand side represents the propagation of the atoms in the potential and the right-hand side stands for the collision integral,
\begin{eqnarray}
I[f_s,f_{\overline{s}}]=\int\frac{d^3p_{\overline{s}}}{(2\pi)^3}\int d\Omega\frac{d\sigma}{d\Omega}&&\frac{|\mathbf{p}_s-\mathbf{p}_{\overline{s}}|}{m}[f_sf_{\overline{s}}(1-f'_s)(1-f'_{\overline{s}})\nonumber \\
&&-f'_sf'_{\overline{s}}(1-f_s)(1-f_{\overline{s}})].
\end{eqnarray}
The indices $s$ and $\overline{s}$ label the two colliding atoms, the primed variables refer to quantities after the collision and $\Omega$ is the solid angle between the incoming and outgoing relative momenta.

In order to simulate the Boltzmann equation we introduce a discrete time step. During each time step the atoms propagate following their classical trajectories. At the end of each time step collisions between the atoms are evaluated. The point-like atom picture is a discrete approximation of the continuous distribution function $f_s(\mathbf{r},\mathbf{p},t)$. In order for this approximation to be accurate we represent each fermion by several test particles \cite{urban, jackson}. The higher the ratio $\tilde{N}/N$ of test particles to atoms, the more precisely the continuous distribution will be approximated. In this work we use $\tilde{N}=10N$, which is sufficient for the range of parameters considered.

The trajectories of the test particles are the same as the trajectories of the atoms and are given by the solution to the  classical harmonic oscillator equations. We evaluate collisions in the same way as described in \cite{urban}. First we test whether a pair of test particles fulfills the classical conditions for scattering. If this is the case we calculate the quantum mechanical scattering probability given by the Pauli term $(1-f'_s)(1-f'_{\overline{s}})$ and accept or reject the collision accordingly. If a collision is accepted we calculate the new phase space coordinates of the particles respecting angular momentum conservation \cite{gueryodelin}.

We performed several tests of the simulation. We ensured that the system thermalises to the correct equilibrium distribution (the Fermi-Dirac distribution in the presence of Pauli blocking, or the Maxwell-Boltzmann distribution if the quantum mechanical scattering probability is set to one). We also checked the frequencies and the damping behavior of several collective modes, namely the dipole, the breathing and the quadrupole mode. A very important check is that the equilibrium collision rate measured during a simulation matches the theoretical prediction, as this will be crucial to obtain the correct damping behavior for the oscillation of two colliding clouds. We performed tests both with and without Pauli blocking and obtained very good agreement with the values obtained analytically from the Fermi-Dirac distribution.

The behavior of the clouds during the simulation can be studied by measuring the distance between their centers of mass $d(t)=\langle z_\uparrow-z_\downarrow \rangle(t)$. As in \cite{Zwierlein}, we find that before they come to rest at thermal equilibrium, the motion of the clouds exhibits three typical behaviors, see Fig.~\ref{figureregimes} and \footnote{See Supplemental Material for an animation showing the time-evolution of the density profiles in the three regimes.}. {\it Transmission:} for sufficiently high temperatures and small interactions, the clouds oscillate through each other (i.e.\ $d(t)$ crosses zero at short times) with decreasing amplitude. {\it Bounce:} at low temperatures and strong interactions the clouds bounce off each other several times (in each bounce the motion of the center of mass of each cloud is reversed at short times and without $d(t)$ crossing zero) before a longer period of slow approach. {\it Intermediate:} between the transmission and bounce regimes there is a range of temperatures and interactions where the slow approach behavior is visible from the start and neither bounces nor transmissions are observed.

\begin{figure*}
\includegraphics[width=0.65\columnwidth]{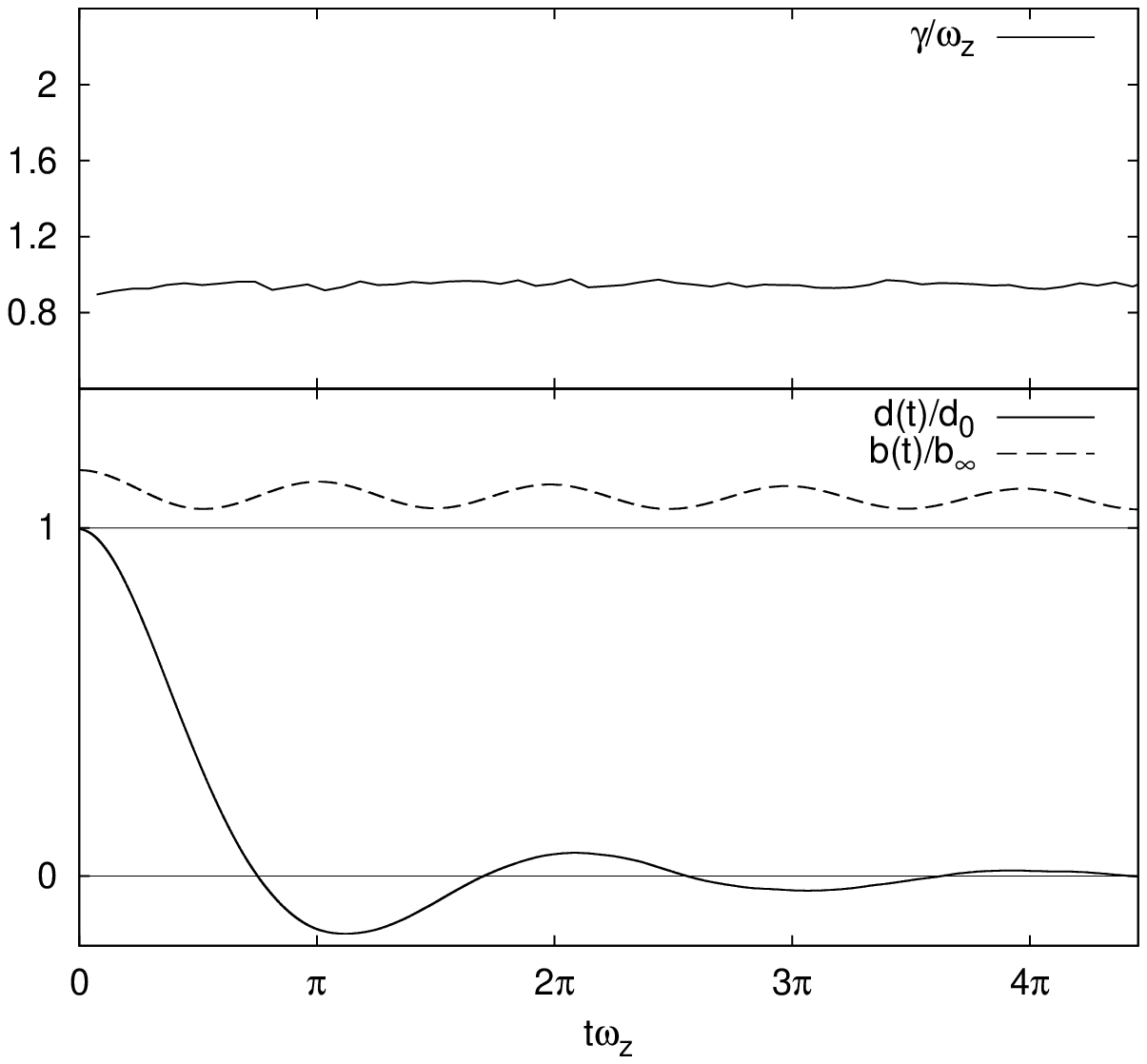}
\hfill
\includegraphics[width=0.65\columnwidth]{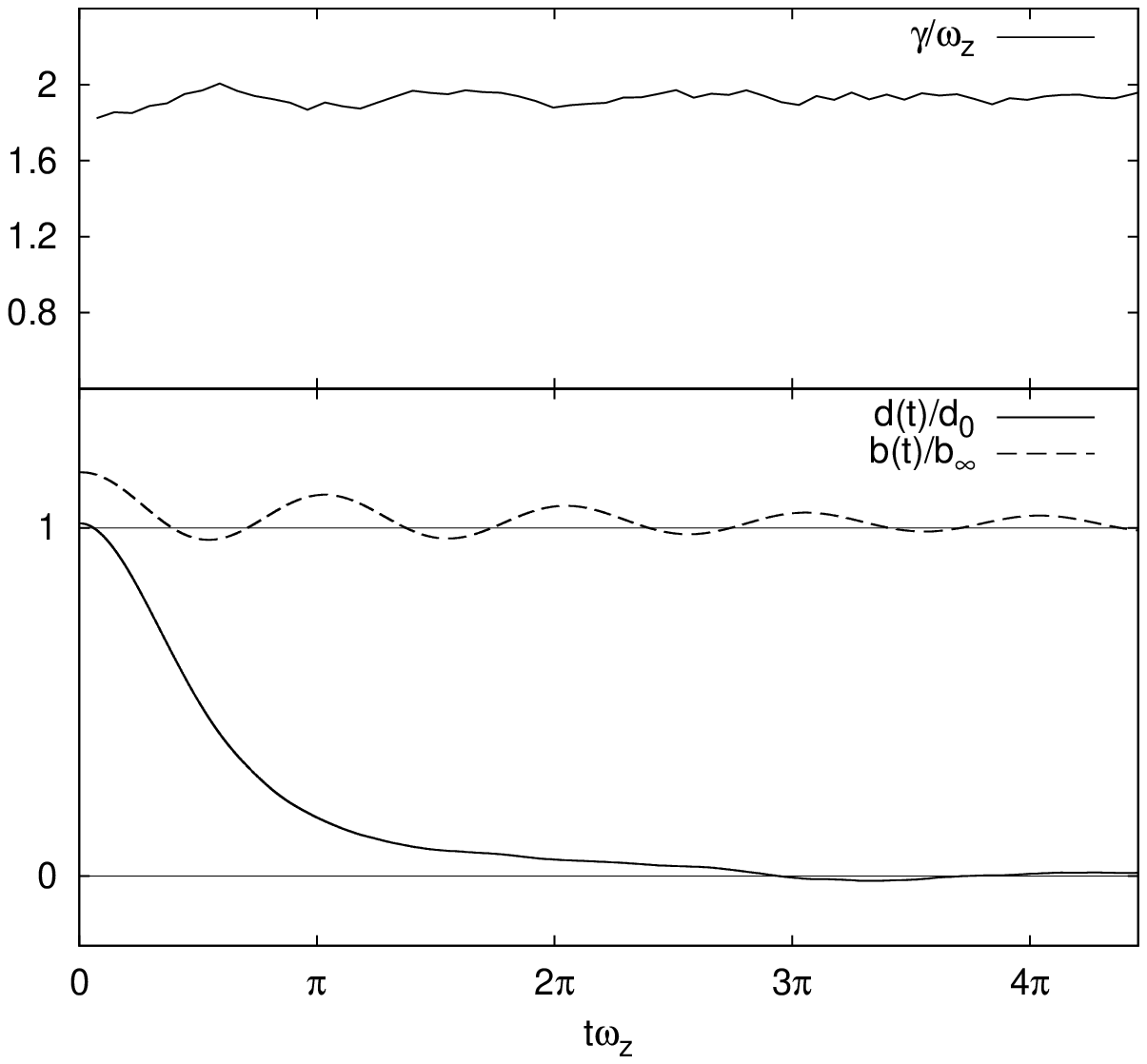}
\hfill
\includegraphics[width=0.65\columnwidth]{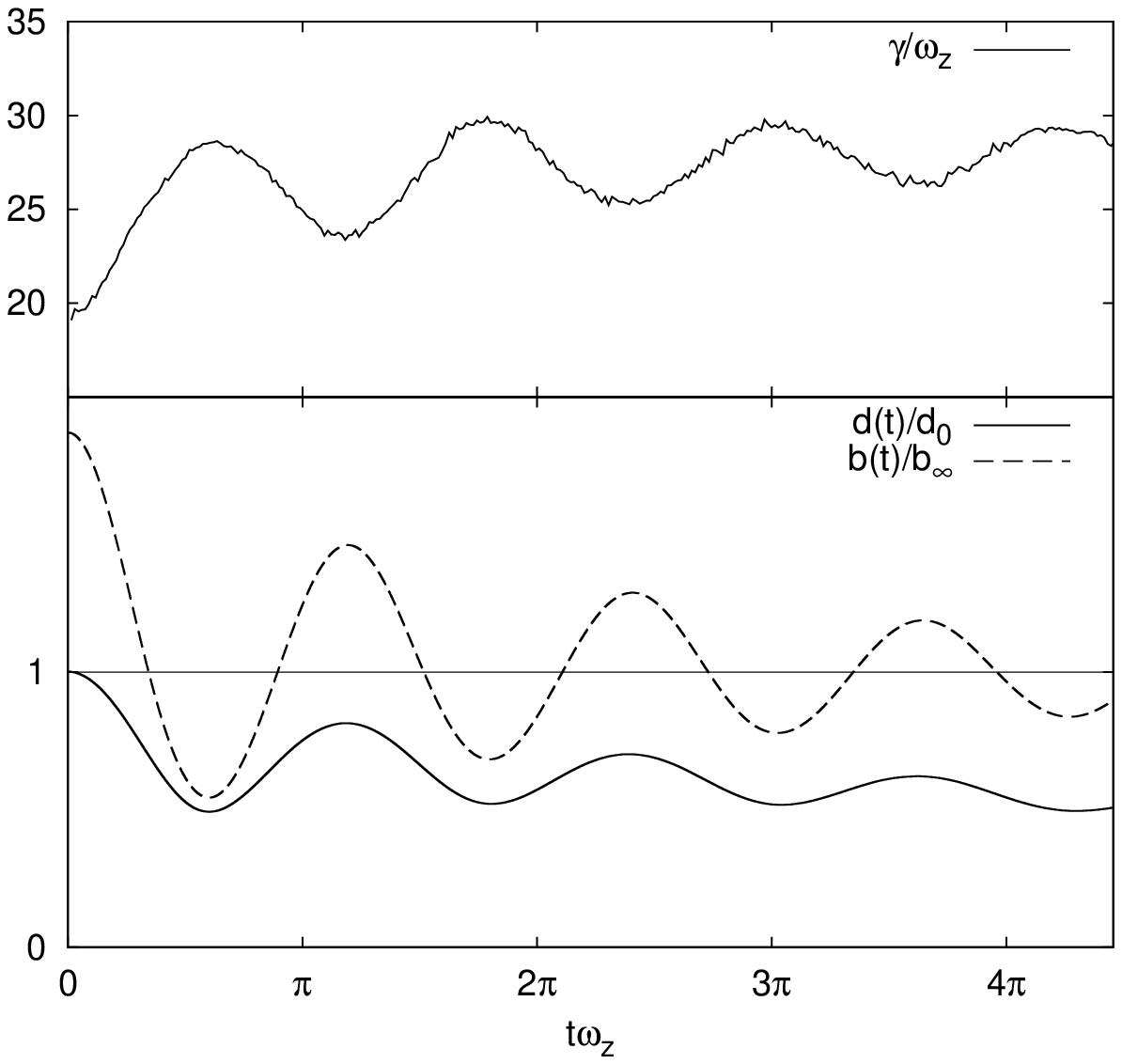}
\caption{Bottom panel: the normalized dipole mode $d(t)/d_0$ (solid lines) and breathing mode $b(t)/b_\infty$ (dashed lines) for the three different behaviors: transmission (left), intermediate (middle) and bounce (right). Top panel: the corresponding collision rate per particle $\gamma/\omega_z$ measured in the region with $|z|\leq\sigma_z/2$ around the trap center, where $\sigma_z=\sqrt{T_\textnormal{init}/m\omega_z^2}$.}
\label{figureregimes}
\end{figure*}
The dependence of the behavior on temperature and interactions is related to the variation in the collision rate $\gamma$ in the overlap region between the two clouds. As the collision rate decreases the system behavior changes from the bounce regime to intermediate, and finally to the transmission regime. From the top panel of Fig.~\ref{figureregimes} we see that in the bounce regime the oscillations in the collision rate integrated over a volume in the overlap region follow closely the oscillations of $d(t)$, whereas no such variation is apparent in the transmission regime. In addition, we can compare the collision rate with the typical timescale for macroscopic motion ($\omega_z^{-1}$). We see that the gas is strongly hydrodynamic in the bounce regime ($\omega_z/\gamma \ll1$) and becomes collisionless in the transmission regime ($\omega_z/\gamma \sim 1$).

\begin{figure}
\includegraphics[width=\columnwidth]{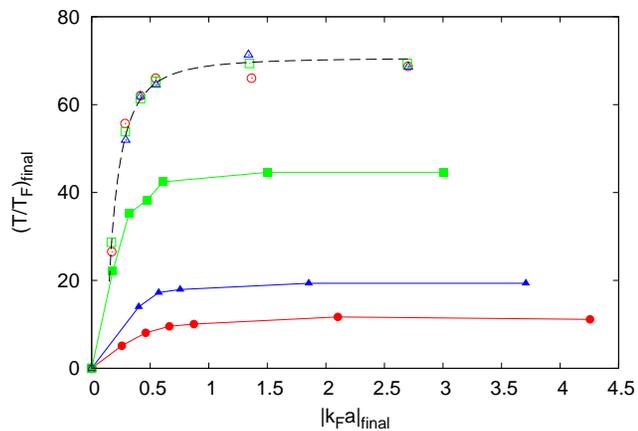}
\caption{\label{fig1}(Color online) The transition between bounce and intermediate regimes (filled symbols, lines to guide the eye) and between intermediate and transmission regimes (empty symbols). Red circles correspond to $d_0=43.1l_z$, blue triangles to  $d_0=64.6l_z$ and green squares to $d_0=129.3l_z$, where $l_z=1/\sqrt{m\omega_z}$. It is clearly visible that the intermediate-transmission transition is independent of $d_0$. The dashed line corresponds to constant relaxation time $1/\tau_{\rm dip}=1.83\omega_z$.}
\end{figure}
Figure \ref{fig1} shows the transitions between these regimes in the $(T/T_F)_\textnormal{final}$, $|k_Fa|_\textnormal{final}$ plane for a fixed aspect ratio and different values of $d_0$ in units $l_z=1/\sqrt{m\omega_z}$. Here the Fermi temperature is defined as $T_F=k_F^2/2m=(3\pi^2n_0)^{2/3}/2m$, where $n_0$ is the total atomic density $n_0=n_{\uparrow 0}+n_{\downarrow 0}$ in the trap center. Note that as the density is a function of the temperature, $T_F$ and $k_F$ change during the simulation and hence $k_Fa$ varies during the evolution. The quantities given here are equilibrium values for $t\rightarrow\infty$.

The transition between transmission and intermediate regimes is defined as $d(t)$ reaching but not crossing zero at short times. It is clearly visible from Fig.~\ref{fig1} that this transition is independent of $d_0$ and hence can be understood entirely from the final equilibrium properties of the system. More precisely, it can be understood as a consequence of the change in the relaxation time $\tau_{\rm dip}$ of the spin dipole mode of the $d_0=0$ system, which is closely related to the collision rate per atom. For sufficiently high temperatures $T\gtrsim \tilde{T}_F$ the relaxation time can be calculated from the Maxwell-Boltzmann distribution \cite{trombettoni} and equals
\begin{equation}
\frac{1}{\tau_{\rm dip}}=\frac{2N}{3\pi T^2}\omega^3f\left(\frac{1}{a^2T}\right), \label{taudipole}
\end{equation}
where $f(y)=1-y+y^2e^y\Gamma(0,y)$. For sufficiently low collision rate ($\omega_z\tau_{\rm dip}\gg 1$), the gas can be said to be collisionless and therefore the clouds undergo independent oscillations without interacting strongly with each other. For $1/\tau_{\rm dip}>2\omega_z$ the dipole mode is overdamped \cite{vichi1999collective, trombettoni}. To compare with the simulations, we calculate the value of $\tau_{\rm dip}$ for the various points lying on the curve separating the transmission and intermediate regimes of Fig.~\ref{fig1} using Eq.~(\ref{taudipole}). We find that they lie on the curve of constant $1/\tau_{\rm dip}=1.83(1)\omega_z$.

The transition between the intermediate and the bounce regime is defined to occur when the first bounce ceases to reverse the motion of the clouds, or in other words when $d(t)$ ceases to have a minimum and becomes a monotonically decreasing function of $t$. This transition depends on $d_0$. In the bounce regime we typically see an initial strong collision followed by oscillations of $d(t)$ which eventually die out as $d(t)\rightarrow0$. This oscillatory behavior continues into the intermediate regime. The bottom panel of Fig.~\ref{figureregimes} shows plots in the three regimes of the normalized amplitudes of the dipole mode $d(t)/d_0$ and the breathing mode $b(t)/b_\infty$ where $b(t)=\langle z^2_\uparrow+z^2_\downarrow\rangle(t)$. In the {\em bounce} regime, the frequency of $d(t)$ is identical to the frequency of $b(t)$, see Fig.~\ref{figureregimes} (right), and suggests the existence of a nonlinear coupling between the two modes. However, in the {\em transmission} regime, see Fig.~\ref{figureregimes} (left), the oscillation frequency of $d(t)$ becomes closer to that of the dipole mode of the non-interacting gas $\omega_z$, and so the spin dipole mode decouples from the breathing mode.

For a more quantitative analysis we fit to the function
\begin{equation}
d(t)=B e^{-t/C}\left(1+D e^{-t/E}\sin\left(\omega t+\phi\right)\right).
\label{fitdipole}
\end{equation}
The first term is related to the spin-drag coefficient measured in \cite{Zwierlein}. It dominates the overdamped behavior of $d(t)$ at long times with a characteristic timescale $C$ which we will analyze elsewhere \cite{inprep}. The second term describes the coupling between the spin dipole and the breathing mode.

The oscillations of the axial size of the cloud are fitted using the dependence
\begin{equation}
b(t)/b_\infty=1+D'e^{-t/E'}\sin\left(\omega' t+\phi'\right).
\label{fitbreathing}
\end{equation}
From the fit to our simulations, we observe that $D'\simeq 2D$, $E'\simeq E$, $\omega'\simeq\omega$ and $\phi'\simeq\phi$. We interpret this result by describing the density profile of the gas by the ansatz $n_s(z,t)=\alpha n_s(\alpha z\pm\beta,0)$, where $\alpha=\alpha(t)$ and $\beta=\beta(t)$ represent the breathing and spin dipole modes respectively. Using this expression, we can calculate $b(t)$ and $d(t)$ as functions of $\alpha$ and $\beta$ and  in the weakly non-linear regime, we obtain $d(t)\simeq 2\beta/\alpha$ and $b(t)\simeq 1/\alpha^2$. Assuming that $\beta(t)$ is overdamped and $\alpha(t)$ is a damped oscillator we then obtain the time dependence in Eqs.~(\ref{fitdipole}) and (\ref{fitbreathing}) and the corresponding relations between the fit coefficients.

\begin{figure}
\includegraphics[width=\columnwidth]{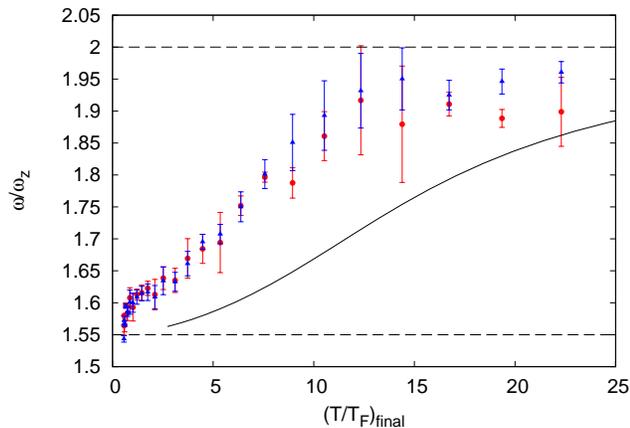}
\caption{\label{frequencyplot}(Color online) The frequency $\omega/\omega_z$ of the dipole mode $d(t)$ (red circles) and the breathing mode $b(t)$ (blue triangles) versus the final temperature for $|\tilde{k}_Fa|=1$. All data were obtained for equal initial temperature $T_{\rm init}=0.4\tilde{T}_F$ by varying $d_0$. The solid line is the prediction from \cite{guery1999collective}.}
\end{figure}
We are also able to study the temperature dependence of the frequency of the two modes, see Fig.~\ref{frequencyplot}. We obtain the frequency by fitting to the functions (\ref{fitdipole}) for the dipole mode and (\ref{fitbreathing}) for the breathing mode. Since the frequency shows a weak time dependence we ignore early times by imposing a cut-off $d_c=2\cdot0.7\sigma_z$ on the amplitude and fitting only times $t>t_c$ for which $d(t)<d_c$. Since the function $d(t)$ is not monotonic the corresponding time cut-off $t_c$ is not continuous with changing $d_0$, which leads to a small systematic error. We estimate this error from the cut-off dependence of $\omega$. The statistical error of the fit is several orders of magnitude smaller and hence negligible.

At low temperature, the common frequency of the spin dipole and breathing modes is close to  the hydrodynamic prediction $\sqrt{12/5}\omega_z\approx1.55\omega_z$ \cite{guery1999collective, amoruso1999collective}, as observed experimentally \cite{Zwierlein}. As we increase temperature, the frequency approaches $2\omega_z$, the non-interacting value. At higher temperatures the spin dipole mode frequency becomes ill-defined due to large damping. If we continue to increase the temperature, the damping becomes progressively smaller and the dipole mode frequency approaches $\omega_z$, the value for the non-interacting gas. We also compare our results with an earlier prediction for the frequency of the breathing mode \cite{guery1999collective}. The two dependencies are close to each other although some discrepancy remains. We attribute it to the fact that to estimate the collisional rate in the cloud Ref.~\cite{guery1999collective} neglects the Pauli principle and assumes a Gaussian phase-space density.

In conclusion, we have studied the collision of two spin polarized fermionic clouds using a Boltzmann equation simulation. We found various regimes of behavior, characterized the transitions between them as a function of interaction strength and temperature, and related them to the collision rate in the overlap region between the clouds. In particular we showed that the bounces can be explained purely as a semiclassical collisional phenomenon, without the need for more complex many-body effects. We also demonstrated that the bounces are a rare example of nonlinear mode coupling, in which the spin dipole and breathing modes interact. In future we aim to extend this study to other closely related problems such as the collision of clouds with unequal populations \cite{ZwierleinImbalanced}, between clouds of atoms with unequal masses \cite{Trenkwalder}, and to the Fermi liquid regime at $T \ll T_F$, using the Landau-Boltzmann equation.

We thank A.~Sommer, M.~Urban and M.~Zwierlein for useful discussions and the INT at the University of Washington for its hospitality. This work has made use of the resources provided by the Cambridge HPC Facility. O.G.\ is supported by the German Academic Exchange Service (DAAD), EPSRC and the Cambridge European Trust. F.C.\ acknowledges support from ERC (project FERLODIM), R\'egion Ile de France (IFRAF) and Institut Universitaire de France. C.L.\ acknowledges support from the EPSRC through the Advanced Fellowship EP/E053033/1.

\bibliography{Boltzmann}

\begin{thebibliography}{21}%
\makeatletter
\providecommand \@ifxundefined [1]{%
 \@ifx{#1\undefined}
}%
\providecommand \@ifnum [1]{%
 \ifnum #1\expandafter \@firstoftwo
 \else \expandafter \@secondoftwo
 \fi
}%
\providecommand \@ifx [1]{%
 \ifx #1\expandafter \@firstoftwo
 \else \expandafter \@secondoftwo
 \fi
}%
\providecommand \natexlab [1]{#1}%
\providecommand \enquote  [1]{``#1''}%
\providecommand \bibnamefont  [1]{#1}%
\providecommand \bibfnamefont [1]{#1}%
\providecommand \citenamefont [1]{#1}%
\providecommand \href@noop [0]{\@secondoftwo}%
\providecommand \href [0]{\begingroup \@sanitize@url \@href}%
\providecommand \@href[1]{\@@startlink{#1}\@@href}%
\providecommand \@@href[1]{\endgroup#1\@@endlink}%
\providecommand \@sanitize@url [0]{\catcode `\\12\catcode `\$12\catcode
  `\&12\catcode `\#12\catcode `\^12\catcode `\_12\catcode `\%12\relax}%
\providecommand \@@startlink[1]{}%
\providecommand \@@endlink[0]{}%
\providecommand \url  [0]{\begingroup\@sanitize@url \@url }%
\providecommand \@url [1]{\endgroup\@href {#1}{\urlprefix }}%
\providecommand \urlprefix  [0]{URL }%
\providecommand \Eprint [0]{\href }%
\providecommand \doibase [0]{http://dx.doi.org/}%
\providecommand \selectlanguage [0]{\@gobble}%
\providecommand \bibinfo  [0]{\@secondoftwo}%
\providecommand \bibfield  [0]{\@secondoftwo}%
\providecommand \translation [1]{[#1]}%
\providecommand \BibitemOpen [0]{}%
\providecommand \bibitemStop [0]{}%
\providecommand \bibitemNoStop [0]{.\EOS\space}%
\providecommand \EOS [0]{\spacefactor3000\relax}%
\providecommand \BibitemShut  [1]{\csname bibitem#1\endcsname}%
\let\auto@bib@innerbib\@empty
\bibitem [{\citenamefont {Fuchs}\ \emph {et~al.}(2003)\citenamefont {Fuchs},
  \citenamefont {Gangardt},\ and\ \citenamefont {Lalo\"e}}]{Bose1}%
  \BibitemOpen
  \bibfield  {author} {\bibinfo {author} {\bibfnamefont {J.~N.}\ \bibnamefont
  {Fuchs}}, \bibinfo {author} {\bibfnamefont {D.~M.}\ \bibnamefont {Gangardt}},
  \ and\ \bibinfo {author} {\bibfnamefont {F.}~\bibnamefont {Lalo\"e}},\ }\href
  {\doibase 10.1140/epjd/e2003-00219-1} {\bibfield  {journal} {\bibinfo
  {journal} {Eur.~Phys.~J.~D}\ }\textbf {\bibinfo {volume} {25}},\ \bibinfo
  {pages} {57} (\bibinfo {year} {2003})}\BibitemShut {NoStop}%
\bibitem [{\citenamefont {McGuirk}(2010)}]{Bose2}%
  \BibitemOpen
  \bibfield  {author} {\bibinfo {author} {\bibfnamefont {J.~M.}\ \bibnamefont
  {McGuirk}},\ }\href {\doibase 10.1103/PhysRevA.82.011612} {\bibfield
  {journal} {\bibinfo  {journal} {Phys.~Rev.~A}\ }\textbf {\bibinfo {volume}
  {82}},\ \bibinfo {pages} {011612} (\bibinfo {year} {2010})}\BibitemShut
  {NoStop}%
\bibitem [{\citenamefont {Du}\ \emph {et~al.}(2008)\citenamefont {Du},
  \citenamefont {Luo}, \citenamefont {Clancy},\ and\ \citenamefont
  {Thomas}}]{du2008observation}%
  \BibitemOpen
  \bibfield  {author} {\bibinfo {author} {\bibfnamefont {X.}~\bibnamefont
  {Du}}, \bibinfo {author} {\bibfnamefont {L.}~\bibnamefont {Luo}}, \bibinfo
  {author} {\bibfnamefont {B.}~\bibnamefont {Clancy}}, \ and\ \bibinfo {author}
  {\bibfnamefont {J.~E.}\ \bibnamefont {Thomas}},\ }\href {\doibase
  10.1103/PhysRevLett.101.150401} {\bibfield  {journal} {\bibinfo  {journal}
  {Phys.~Rev.~Lett.}\ }\textbf {\bibinfo {volume} {101}},\ \bibinfo {pages}
  {150401} (\bibinfo {year} {2008})}\BibitemShut {NoStop}%
\bibitem [{\citenamefont {Du}\ \emph {et~al.}(2009)\citenamefont {Du},
  \citenamefont {Zhang}, \citenamefont {Petricka},\ and\ \citenamefont
  {Thomas}}]{du2009controlling}%
  \BibitemOpen
  \bibfield  {author} {\bibinfo {author} {\bibfnamefont {X.}~\bibnamefont
  {Du}}, \bibinfo {author} {\bibfnamefont {Y.}~\bibnamefont {Zhang}}, \bibinfo
  {author} {\bibfnamefont {J.}~\bibnamefont {Petricka}}, \ and\ \bibinfo
  {author} {\bibfnamefont {J.~E.}\ \bibnamefont {Thomas}},\ }\href {\doibase
  10.1103/PhysRevLett.103.010401} {\bibfield  {journal} {\bibinfo  {journal}
  {Phys.~Rev.~Lett.}\ }\textbf {\bibinfo {volume} {103}},\ \bibinfo {pages}
  {010401} (\bibinfo {year} {2009})}\BibitemShut {NoStop}%
\bibitem [{\citenamefont {Pi\'echon}\ \emph {et~al.}(2009)\citenamefont
  {Pi\'echon}, \citenamefont {Fuchs},\ and\ \citenamefont
  {Lalo\"e}}]{piechon2009large}%
  \BibitemOpen
  \bibfield  {author} {\bibinfo {author} {\bibfnamefont {F.}~\bibnamefont
  {Pi\'echon}}, \bibinfo {author} {\bibfnamefont {J.~N.}\ \bibnamefont
  {Fuchs}}, \ and\ \bibinfo {author} {\bibfnamefont {F.}~\bibnamefont
  {Lalo\"e}},\ }\href {\doibase 10.1103/PhysRevLett.102.215301} {\bibfield
  {journal} {\bibinfo  {journal} {Phys.~Rev.~Lett.}\ }\textbf {\bibinfo
  {volume} {102}},\ \bibinfo {pages} {215301} (\bibinfo {year}
  {2009})}\BibitemShut {NoStop}%
\bibitem [{\citenamefont {{Duine}}\ \emph {et~al.}(2011)\citenamefont
  {{Duine}}, \citenamefont {{Polini}}, \citenamefont {{Raoux}}, \citenamefont
  {{Stoof}},\ and\ \citenamefont {{Vignale}}}]{Duine}%
  \BibitemOpen
  \bibfield  {author} {\bibinfo {author} {\bibfnamefont {R.~A.}\ \bibnamefont
  {{Duine}}}, \bibinfo {author} {\bibfnamefont {M.}~\bibnamefont {{Polini}}},
  \bibinfo {author} {\bibfnamefont {A.}~\bibnamefont {{Raoux}}}, \bibinfo
  {author} {\bibfnamefont {H.~T.~C.}\ \bibnamefont {{Stoof}}}, \ and\ \bibinfo
  {author} {\bibfnamefont {G.}~\bibnamefont {{Vignale}}},\ }\href {\doibase
  10.1088/1367-2630/13/4/045010} {\bibfield  {journal} {\bibinfo  {journal}
  {New~J.~Phys.}\ }\textbf {\bibinfo {volume} {13}},\ \bibinfo {pages} {045010}
  (\bibinfo {year} {2011})}\BibitemShut {NoStop}%
\bibitem [{\citenamefont {{Sommer}}\ \emph
  {et~al.}(2011{\natexlab{a}})\citenamefont {{Sommer}}, \citenamefont {{Ku}},
  \citenamefont {{Roati}},\ and\ \citenamefont {{Zwierlein}}}]{Zwierlein}%
  \BibitemOpen
  \bibfield  {author} {\bibinfo {author} {\bibfnamefont {A.}~\bibnamefont
  {{Sommer}}}, \bibinfo {author} {\bibfnamefont {M.}~\bibnamefont {{Ku}}},
  \bibinfo {author} {\bibfnamefont {G.}~\bibnamefont {{Roati}}}, \ and\
  \bibinfo {author} {\bibfnamefont {M.~W.}\ \bibnamefont {{Zwierlein}}},\
  }\href {\doibase 10.1038/nature09989} {\bibfield  {journal} {\bibinfo
  {journal} {Nature}\ }\textbf {\bibinfo {volume} {472}},\ \bibinfo {pages}
  {201} (\bibinfo {year} {2011}{\natexlab{a}})}\BibitemShut {NoStop}%
\bibitem [{\citenamefont {{Bruun}}(2011)}]{Bruun}%
  \BibitemOpen
  \bibfield  {author} {\bibinfo {author} {\bibfnamefont {G.~M.}\ \bibnamefont
  {{Bruun}}},\ }\href {\doibase 10.1088/1367-2630/13/3/035005} {\bibfield
  {journal} {\bibinfo  {journal} {New~J.~Phys.}\ }\textbf {\bibinfo {volume}
  {13}},\ \bibinfo {pages} {035005} (\bibinfo {year} {2011})}\BibitemShut
  {NoStop}%
\bibitem [{\citenamefont {{Taylor}}\ \emph {et~al.}(2011)\citenamefont
  {{Taylor}}, \citenamefont {{Zhang}}, \citenamefont {{Schneider}},\ and\
  \citenamefont {{Randeria}}}]{Taylor2011Colliding}%
  \BibitemOpen
  \bibfield  {author} {\bibinfo {author} {\bibfnamefont {E.}~\bibnamefont
  {{Taylor}}}, \bibinfo {author} {\bibfnamefont {S.}~\bibnamefont {{Zhang}}},
  \bibinfo {author} {\bibfnamefont {W.}~\bibnamefont {{Schneider}}}, \ and\
  \bibinfo {author} {\bibfnamefont {M.}~\bibnamefont {{Randeria}}},\
  }\href@noop {} {\  (\bibinfo {year} {2011})},\ \Eprint
  {http://arxiv.org/abs/1106.4245} {arXiv:1106.4245 [cond-mat.quant-gas]}
  \BibitemShut {NoStop}%
\bibitem [{Note1()}]{Note1}%
  \BibitemOpen
  \bibinfo {note} {In Bose gases, nonlinear coupling between modes can lead to
  damping of collective excitations, as in the case of Landau and Beliaev
  damping.}\BibitemShut {Stop}%
\bibitem [{\citenamefont {Lepers}\ \emph {et~al.}(2010)\citenamefont {Lepers},
  \citenamefont {Davesne}, \citenamefont {Chiacchiera},\ and\ \citenamefont
  {Urban}}]{urban}%
  \BibitemOpen
  \bibfield  {author} {\bibinfo {author} {\bibfnamefont {T.}~\bibnamefont
  {Lepers}}, \bibinfo {author} {\bibfnamefont {D.}~\bibnamefont {Davesne}},
  \bibinfo {author} {\bibfnamefont {S.}~\bibnamefont {Chiacchiera}}, \ and\
  \bibinfo {author} {\bibfnamefont {M.}~\bibnamefont {Urban}},\ }\href
  {\doibase 10.1103/PhysRevA.82.023609} {\bibfield  {journal} {\bibinfo
  {journal} {Phys.~Rev.~A}\ }\textbf {\bibinfo {volume} {82}},\ \bibinfo
  {pages} {023609} (\bibinfo {year} {2010})}\BibitemShut {NoStop}%
\bibitem [{\citenamefont {Goulko}\ \emph {et~al.}()\citenamefont {Goulko},
  \citenamefont {Chevy},\ and\ \citenamefont {Lobo}}]{inprep}%
  \BibitemOpen
  \bibfield  {author} {\bibinfo {author} {\bibfnamefont {O.}~\bibnamefont
  {Goulko}}, \bibinfo {author} {\bibfnamefont {F.}~\bibnamefont {Chevy}}, \
  and\ \bibinfo {author} {\bibfnamefont {C.}~\bibnamefont {Lobo}},\ }\href@noop
  {} {}\bibinfo {note} {In preparation}\BibitemShut {NoStop}%
\bibitem [{\citenamefont {Jackson}\ and\ \citenamefont
  {Zaremba}(2002)}]{jackson}%
  \BibitemOpen
  \bibfield  {author} {\bibinfo {author} {\bibfnamefont {B.}~\bibnamefont
  {Jackson}}\ and\ \bibinfo {author} {\bibfnamefont {E.}~\bibnamefont
  {Zaremba}},\ }\href {\doibase 10.1103/PhysRevA.66.033606} {\bibfield
  {journal} {\bibinfo  {journal} {Phys.~Rev.~A}\ }\textbf {\bibinfo {volume}
  {66}},\ \bibinfo {pages} {033606} (\bibinfo {year} {2002})}\BibitemShut
  {NoStop}%
\bibitem [{\citenamefont {Gu\'ery-Odelin}(1998)}]{gueryodelin}%
  \BibitemOpen
  \bibfield  {author} {\bibinfo {author} {\bibfnamefont {D.}~\bibnamefont
  {Gu\'ery-Odelin}},\ }\emph {\bibinfo {title} {{Dynamique collisionnelle des
  gaz d'alcalins lourds: du refroidissement evaporatif a la condensation de
  Bose-Einstein}}},\ \href@noop {} {Ph.D. thesis},\ \bibinfo  {school}
  {Universit\'e Paris VI} (\bibinfo {year} {1998})\BibitemShut {NoStop}%
\bibitem [{Note2()}]{Note2}%
  \BibitemOpen
  \bibinfo {note} {See Supplemental Material for an animation showing the
  time-evolution of the density profiles in the three regimes.}\BibitemShut
  {Stop}%
\bibitem [{\citenamefont {Chiacchiera}\ \emph {et~al.}(2010)\citenamefont
  {Chiacchiera}, \citenamefont {Macr\`\i{}},\ and\ \citenamefont
  {Trombettoni}}]{trombettoni}%
  \BibitemOpen
  \bibfield  {author} {\bibinfo {author} {\bibfnamefont {S.}~\bibnamefont
  {Chiacchiera}}, \bibinfo {author} {\bibfnamefont {T.}~\bibnamefont
  {Macr\`\i{}}}, \ and\ \bibinfo {author} {\bibfnamefont {A.}~\bibnamefont
  {Trombettoni}},\ }\href {\doibase 10.1103/PhysRevA.81.033624} {\bibfield
  {journal} {\bibinfo  {journal} {Phys.~Rev.~A}\ }\textbf {\bibinfo {volume}
  {81}},\ \bibinfo {pages} {033624} (\bibinfo {year} {2010})}\BibitemShut
  {NoStop}%
\bibitem [{\citenamefont {Vichi}\ and\ \citenamefont
  {Stringari}(1999)}]{vichi1999collective}%
  \BibitemOpen
  \bibfield  {author} {\bibinfo {author} {\bibfnamefont {L.}~\bibnamefont
  {Vichi}}\ and\ \bibinfo {author} {\bibfnamefont {S.}~\bibnamefont
  {Stringari}},\ }\href {\doibase 10.1103/PhysRevA.60.4734} {\bibfield
  {journal} {\bibinfo  {journal} {Phys.~Rev.~A}\ }\textbf {\bibinfo {volume}
  {60}},\ \bibinfo {pages} {4734} (\bibinfo {year} {1999})}\BibitemShut
  {NoStop}%
\bibitem [{\citenamefont {Gu\'ery-Odelin}\ \emph {et~al.}(1999)\citenamefont
  {Gu\'ery-Odelin}, \citenamefont {Zambelli}, \citenamefont {Dalibard},\ and\
  \citenamefont {Stringari}}]{guery1999collective}%
  \BibitemOpen
  \bibfield  {author} {\bibinfo {author} {\bibfnamefont {D.}~\bibnamefont
  {Gu\'ery-Odelin}}, \bibinfo {author} {\bibfnamefont {F.}~\bibnamefont
  {Zambelli}}, \bibinfo {author} {\bibfnamefont {J.}~\bibnamefont {Dalibard}},
  \ and\ \bibinfo {author} {\bibfnamefont {S.}~\bibnamefont {Stringari}},\
  }\href {\doibase 10.1103/PhysRevA.60.4851} {\bibfield  {journal} {\bibinfo
  {journal} {Phys.~Rev.~A}\ }\textbf {\bibinfo {volume} {60}},\ \bibinfo
  {pages} {4851} (\bibinfo {year} {1999})}\BibitemShut {NoStop}%
\bibitem [{\citenamefont {Amoruso}\ \emph {et~al.}(1999)\citenamefont
  {Amoruso}, \citenamefont {Meccoli}, \citenamefont {Minguzzi},\ and\
  \citenamefont {Tosi}}]{amoruso1999collective}%
  \BibitemOpen
  \bibfield  {author} {\bibinfo {author} {\bibfnamefont {M.}~\bibnamefont
  {Amoruso}}, \bibinfo {author} {\bibfnamefont {I.}~\bibnamefont {Meccoli}},
  \bibinfo {author} {\bibfnamefont {A.}~\bibnamefont {Minguzzi}}, \ and\
  \bibinfo {author} {\bibfnamefont {M.~P.}\ \bibnamefont {Tosi}},\ }\href
  {\doibase 10.1007/s100530050588} {\bibfield  {journal} {\bibinfo  {journal}
  {Eur.~Phys.~J.~D}\ }\textbf {\bibinfo {volume} {7}},\ \bibinfo {pages} {441}
  (\bibinfo {year} {1999})}\BibitemShut {NoStop}%
\bibitem [{\citenamefont {{Sommer}}\ \emph
  {et~al.}(2011{\natexlab{b}})\citenamefont {{Sommer}}, \citenamefont {{Ku}},\
  and\ \citenamefont {{Zwierlein}}}]{ZwierleinImbalanced}%
  \BibitemOpen
  \bibfield  {author} {\bibinfo {author} {\bibfnamefont {A.}~\bibnamefont
  {{Sommer}}}, \bibinfo {author} {\bibfnamefont {M.}~\bibnamefont {{Ku}}}, \
  and\ \bibinfo {author} {\bibfnamefont {M.~W.}\ \bibnamefont {{Zwierlein}}},\
  }\href {\doibase 10.1088/1367-2630/13/5/055009} {\bibfield  {journal}
  {\bibinfo  {journal} {New~J.~Phys.}\ }\textbf {\bibinfo {volume} {13}},\
  \bibinfo {pages} {055009} (\bibinfo {year} {2011}{\natexlab{b}})}\BibitemShut
  {NoStop}%
\bibitem [{\citenamefont {Trenkwalder}\ \emph {et~al.}(2011)\citenamefont
  {Trenkwalder}, \citenamefont {Kohstall}, \citenamefont {Zaccanti},
  \citenamefont {Naik}, \citenamefont {Sidorov}, \citenamefont {Schreck},\ and\
  \citenamefont {Grimm}}]{Trenkwalder}%
  \BibitemOpen
  \bibfield  {author} {\bibinfo {author} {\bibfnamefont {A.}~\bibnamefont
  {Trenkwalder}}, \bibinfo {author} {\bibfnamefont {C.}~\bibnamefont
  {Kohstall}}, \bibinfo {author} {\bibfnamefont {M.}~\bibnamefont {Zaccanti}},
  \bibinfo {author} {\bibfnamefont {D.}~\bibnamefont {Naik}}, \bibinfo {author}
  {\bibfnamefont {A.~I.}\ \bibnamefont {Sidorov}}, \bibinfo {author}
  {\bibfnamefont {F.}~\bibnamefont {Schreck}}, \ and\ \bibinfo {author}
  {\bibfnamefont {R.}~\bibnamefont {Grimm}},\ }\href {\doibase
  10.1103/PhysRevLett.106.115304} {\bibfield  {journal} {\bibinfo  {journal}
  {Phys.~Rev.~Lett.}\ }\textbf {\bibinfo {volume} {106}},\ \bibinfo {pages}
  {115304} (\bibinfo {year} {2011})}\BibitemShut {NoStop}%
\end{thebibliography}%

\end{document}